\documentclass[comment, prl,twocolumn,nofootinbib, preprintnumbers, superscriptaddress]{revtex4-2}

\usepackage{amsmath,amssymb,slashed,braket}
\usepackage{graphicx}
\usepackage{cancel}
\usepackage{epstopdf}
\usepackage{float}
\usepackage[colorlinks=true,
            linkcolor=blue,
            urlcolor=blue,
            citecolor=green,          
            bookmarks=true,
            bookmarksnumbered=true,
            breaklinks=true,
            pdfpagemode=Fullscreen,
            pdfstartview=FitBH]{hyperref}

\usepackage[normalem]{ulem}
\usepackage{color}

\definecolor{Orange}{cmyk}{0,0.61,0.87,0}
\definecolor{JungleGreen}{cmyk}{0.99,0,0.52,0}
\definecolor{OliveGreen}{cmyk}{0.64,0,0.95,0.40}
\definecolor{Brown}{cmyk}{0,0.81,1,0.60}
\definecolor{RoyalBlue}{cmyk}{0.71,0.53,0,0.12}
\definecolor{Gray}{cmyk}{0,0,0,0.40}
\definecolor{LightPink}{cmyk}{0.0,0.25,0,0}
\definecolor{LLightPink}{cmyk}{0.0,0.10,0,0}
\definecolor{LightBlue}{cmyk}{0.25,0,0,0}
\definecolor{LightGray}{cmyk}{0,0,0,0.2}

\usepackage{xcolor}
\definecolor{gesfpurple}{rgb}{0.47,0.19,0.42}

\definecolor{gesflanse}{rgb}{0.00,0.50,0.50}

\definecolor{gesfblue}{rgb}{0.08,0.42,0.76}

\definecolor{gesfred}{rgb}{1,0,0}

\definecolor{gesfwhite}{rgb}{1,1,1}

\definecolor{gesfblack}{rgb}{0,0,0}

\newcommand{\geqn}[1]{Eq.\,\hypersetup{linkcolor=blue}(\ref{#1})\hypersetup{linkcolor=blue}}
\newcommand{\gfig}[1]{{\hypersetup{linkcolor=violet}Fig.\,\ref{#1}\hypersetup{linkcolor=blue}}}

\graphicspath{{Figures/}}

\begin{document}

\title{
Testing the Origins of Neutrino Mass with Supernova Neutrino Time Delay
}

\author{Shao-Feng Ge}
\email{gesf@sjtu.edu.cn}
\affiliation{Tsung-Dao Lee Institute \& School of Physics and Astronomy, Shanghai Jiao Tong University, China}
\affiliation{Key Laboratory for Particle Astrophysics and Cosmology (MOE) 
\& Shanghai Key Laboratory for Particle Physics and Cosmology, Shanghai Jiao Tong University, Shanghai 200240, China}

\author{Chui-Fan Kong}
\email{kongcf@sjtu.edu.cn}
\affiliation{Tsung-Dao Lee Institute \& School of Physics and Astronomy, Shanghai Jiao Tong University, China}
\affiliation{Key Laboratory for Particle Astrophysics and Cosmology (MOE) 
\& Shanghai Key Laboratory for Particle Physics and Cosmology, Shanghai Jiao Tong University, Shanghai 200240, China}

\author{Alexei Y. Smirnov}
\email{smirnov@mpi-hd.mpg.de}
\affiliation{Max-Planck-Institut für Kernphysik, Saupfercheckweg 1, 69117 Heidelberg, Germany}

\begin{abstract}
The origin of neutrino masses 
remains unknown. Both the vacuum mass and the 
dark mass generated by the neutrino interaction with dark matter (DM) particles or fields can fit the current oscillation data.  
The dark mass squared is proportional to the 
DM number density and therefore varies on 
the galactic scale with much larger values
around the Galactic Center. 
This affects the group velocity and the arrival time delay
of core-collapse supernovae (SN) neutrinos.  
This time delay, especially for the  
$\nu_e$ neutronization peak with a sharp time structure, 
can be used to distinguish the vacuum and dark neutrino masses.
For illustration, we explore the potential 
of DUNE which is sensitive to $\nu_e$.  
Our simulations show that DUNE can 
distinguish the two neutrino mass origins 
at more than $5\sigma$\,C.L., depending on
the observed local value of neutrino mass, the neutrino mass ordering, 
the DM density profile, and the SN location. 
\end{abstract}

\maketitle 

\textbf{Introduction} -- 
Numerous data from atmospheric, reactor, and 
accelerator neutrino experiments provided robust
evidence for neutrino oscillation 
\cite{Pontecorvo:1957cp,Pontecorvo:1967fh,Maki:1962mu,PDG22-NuCP}. 
The MSW (Mikheyev-Smirnov-Wolfenstein) effect 
\cite{Wolfenstein:1977ue,Mikheyev:1985zog},
was established as the solution to the solar neutrino problem 
\cite{Maltoni:2015kca,Smirnov:2016xzf}.
The observed energy dependence of these effects 
is consistent with non-zero 
neutrino masses $m_{i}$ and the terms 
$ \Delta m_{ij}^2 / 2 E_\nu$ 
($\Delta m_{ ij}^2  \equiv  m_{i}^2 - m_{j}^2$) 
in the Hamiltonian of flavor evolution for
ultra-relativistic neutrinos. 

Still, the origin of neutrino oscillations and the
nature of neutrino masses are not fully established. 
Indeed, oscillations are sensitive to the
neutrino masses squared that do not change chirality
and therefore do not probe the masses directly.
The term $A / 2 E_\nu$ of any origin 
in the Hamiltonian of evolution, where $A$ is a constant, 
can reproduce the observed oscillation data. 
In fact, neutrino forward 
scattering on background particles, e.g. on light dark matter 
such as the fuzzy/wavy DM, 
generates a potential with $A / 2 E_\nu$ dependence 
\cite{Ge:2019tdi, Choi:2019zxy,Choi:2020ydp,Sen:2023uga}
and $A$ can be interpreted as the
effective mass squared. 
For brevity, we will call this effective mass the dark  
mass, $A = m_{\rm dark}^2$, to emphasize its origin.

There are several ways to realize such a possibility 
\cite{Ge:2019tdi,Choi:2019zxy,Choi:2020ydp,Sen:2023uga}
which differ by properties of DM particles and
mediators of interactions. 
The dark masses squared  are proportional to the 
number density of DM particles $n_{\rm DM}(x)$
\cite{Ge:2019tdi}, and in general 
are functions of the neutrino energy $E_\nu$ 
and time \cite{Berlin:2016woy,Sen:2023uga}:
\begin{equation}
  m_{\rm dark}^2
=  
  n_{\rm DM}({\mathbf x},t) g(t) f(E_\nu).    
\label{eq:refmass}
\end{equation}
The functions $g(t)$ and $f(E_\nu)$ depend on specific models   
and for simplicity we assume $g(t) = 1$. 

According to \geqn{eq:refmass}, the value of 
dark mass varies with space-time coordinates.  
To explain experimental data,
the DM density and consequently the
dark masses should be almost 
constants in the Solar System. However,
they vary significantly on the galactic 
scales. In this paper, we propose to use the galactic
SN neutrinos to probe the space-time 
dependence and therefore the nature of neutrino masses. 
Specifically, we suggest searching for the excessive time delay and 
spread of the $\nu_e$ neutronization peak that has a 
sharp time structure.

\textbf{Time Delay due to the Dark Neutrino Mass} --
The idea of measuring the neutrino masses
via the arrival time delay of SN neutrinos 
was put forward long time ago by 
Zatsepin \cite{Zatsepin:1968kt}.
The time delay due to the vacuum mass with respect to 
massless particles, $\Delta t_{\rm vac} =  (m_{\rm vac} / E_\nu)^2 (D / 2 )$,
after traveling distance $D$ is
\begin{equation}
  \Delta t_{\rm vac}
\approx 
  5.14\,{\rm ms}
  \left(\frac{m_{\rm vac}}{\rm eV}\right)^2
  \left(\frac{10\, {\rm MeV}}{E_\nu}\right)^2
  \frac{D}{10\,{\rm kpc}}.
\label{eq:time-delay-usual}
\end{equation}
Analyzing the observed neutrino events 
associated with SN1987A, an upper 
bound $m_\nu< 5.8\,$eV (95\%\,C.L.) was established \cite{Loredo:2001rx,Pagliaroli:2010ik}. 
The sensitivity can be improved down to the sub-eV level 
with future galactic SN neutrino observations  
\cite{Nardi:2003pr, Nardi:2004zg, Lu:2014zma, 
Hyper-Kamiokande:2018ofw, Hansen:2019giq, Pompa:2022cxc,Pompa:2023yzg,Parker:2023cos}.

The value of dark mass in \geqn{eq:refmass} changes on the way from a star 
to the Earth. Since the DM number density 
increases substantially towards the Galactic Center,
the SN neutrinos with varying dark mass may have much larger delay  
than in the case of vacuum mass. 
The dispersion relation for neutrinos with dark mass
only can be written as  
$ E_\nu = p_\nu + V$, 
where $p_\nu$ is the neutrino momentum and
$V \equiv m^2_{\rm dark} / 2 p_\nu$ is the effective potential.
Consequently, the group velocity $v \equiv d E_\nu / d p_\nu$
differs from the speed of light by
\begin{equation}
  1 -   v(\mathbf{x}) 
=   
  \frac{m_{\rm dark}^2 (\mathbf{x}, p_\nu)}{2p_\nu^2} 
  - 
  \frac{1}{2 p_\nu} \frac{d m_{\rm dark}^2 (\mathbf{x}, p_\nu) }{dp_\nu}. 
\label{eq:group2}
\end{equation}
Here the first term is similar to that for vacuum mass while the second 
one arises from the possible energy dependence of the dark mass.  
Experimentally, the energy dependence of oscillation 
parameters has not been found for $E_\nu \gtrsim 1$\,MeV 
relevant for SN neutrinos and therefore the last 
term in \geqn{eq:group2} should be small.

The time delay due to the dark mass
is given by the integral over the neutrino trajectory
from a SN at coordinate $\mathbf{x}_\star$
to the Sun  $\mathbf{x}_\odot$
($|\mathbf{x}_\star - \mathbf{x}_\odot| \equiv D$):
$\Delta t_{\rm dark} \equiv \int_{\mathbf{x}_\star}^{\mathbf{x}_\odot}
  [1/v(\mathbf{x}) - 1] |d \mathbf{x}| \approx \int_{\mathbf{x}_\star}^{\mathbf{x}_\odot}
  [1 - v(\mathbf{x})] |d \mathbf{x}|$.

We take for illustration
the study \cite{Sen:2023uga}
in which the DM is a light boson $\phi$ with mass
$m_\phi$ and number density $n_\phi$. 
The interaction of neutrinos with DM arises from the 
effective Lagrangian $ \mathcal{L}_{\rm eff} \supset \lambda \bar{\chi} \nu \phi + h.c.$ \footnote{
The effective coupling $\mathcal{L}_{\rm eff}$ can be generated, e.g., 
via the gauge invariant Lagrangian 
$\mathcal{L} \supset  Y \bar \chi_R \widetilde \Phi^\dagger L   + 
\Lambda \phi H^\dagger \Phi +h.c.  + ...$ \cite{Farzan:2018gtr}, where $L$ and $H$ are the 
Standard Model leptonic and Higgs $SU(2)$ 
doublets while $\Phi$ is a second Higgs doublet with large mass. 
We impose a dark $\mathbb Z_2$ odd parity on the new fields $\chi$, $\Phi$, and $\phi$ to forbid other
Yukawa interactions. The second term of the Lagrangian produces the mixing 
between $\phi$ and $\Phi$: $\sim \Lambda  \langle H \rangle$, and consequently,  
the coupling of $\phi$ with neutrinos. The phenomenology of this model has been extensively discussed   
in \cite{Farzan:2018gtr}.},
where
$\lambda$ is the coupling constant and
$\chi$ is the light fermionic mediator with mass $m_\chi$. 
The elastic forward scattering 
$\nu \phi \rightarrow \nu \phi$ 
through $\chi$ mediation in the $s$- and 
$t$-channels produces the effective potential 
and therefore the dark mass squared of the form 
\geqn{eq:refmass} \cite{Ge:2019tdi,Sen:2023uga} 
\begin{equation}
  m^2_{\rm dark}(\mathbf{x})
=
  \frac{\lambda^2 n_\phi (\mathbf{x})}{m_\phi} f(E_\nu)
=
  \frac{\lambda^2 \rho_\phi (\mathbf{x})}{m^2_\phi} f(E_\nu),
\label{eq:envmass}
\end{equation}
where
$ f(E_\nu) \equiv y (y - \epsilon)/(y^2-1)$,
$y \equiv E_\nu/E_R \approx p_\nu/p_R$.
Here 
$E_R \equiv m_\chi^2/2 m_\phi$ is the resonance energy and
$\epsilon$ is the DM charge asymmetry \cite{Sen:2023uga}.
In \geqn{eq:envmass} the DM energy density equals
$\rho_\phi \approx m_\phi n_\phi$
since DM particles are non-relativistic. 
For $p_\nu \gg p_R$, $f \approx 1$. Let us introduce the dark mass inside the Solar System
$m^2_{\rm dark}(\mathbf{x}_\odot) \equiv \lambda^2
\rho_\phi (\mathbf{x}_\odot)/m^2_\phi$ which
can be measured in the Earth-based neutrino experiments.
Then from \geqn{eq:envmass} and \geqn{eq:group2} we find  
\begin{equation}
  1 - v(\mathbf{x})
\approx
  \frac{\lambda^2}
       {2 E^2_\nu m^2_\phi}
       \rho_\phi(\mathbf{x}) f_2 \approx
\frac{m^2_{\rm dark}(\mathbf{x}_\odot) }{2E_\nu^2}
\frac{\rho_\phi(\mathbf{x})}{\rho_\phi(\mathbf{x}_\odot)} f_2,
\label{eq:devia}
\end{equation}
with
$f_2 \equiv y^2(y^2 - 2 \epsilon y +1)/(y^2 - 1)^2$,
and  $f_2 \approx 1$ for $p_\nu \gg p_R$.
Consequently, the time delay equals
\begin{equation}
\hspace{-3mm}
  \Delta t_{\rm dark}
=
  \frac {m^2_{\rm dark}(\mathbf{x}_\odot)D}{2E_\nu^2}
  \frac{\int_{\mathbf{x}_\star}^{\mathbf{x}_\odot}
  \rho_\phi (\mathbf{x}) |d \mathbf{x}|}{D \rho_\phi (\mathbf{x}_\odot)}
\equiv
  \Delta t_{\odot} \frac{\overline{\rho_\phi}(\mathbf{x}_\star)}{\rho_\phi (\mathbf{x}_\odot)},
\label{eq:time-delay-env2}
\end{equation}
where $\Delta t_{\odot}\equiv m^2_{\rm dark}(\mathbf{x}_\odot)D/2E_\nu^2$
is the time delay due to the local dark mass
$m^2_{\rm dark}(\mathbf{x}_\odot)$.  Here  $\rho_\phi (\mathbf{x}_\odot)$
is the local DM energy density
in the Solar System and
$
\overline{\rho_\phi}
(\mathbf{x}_\star)\equiv
\int_{\mathbf{x}_\star}^{\mathbf{x}_\odot}
\rho_\phi (\mathbf{x}) |d \mathbf{x}|/D
$ is the average DM energy density along the neutrino trajectory.
In addition to modifying the neutrino
oscillation, the effective coupling $\mathcal{L}_{\rm eff}$ also induces
the $\nu\rightarrow \chi$ transition and oscillations in the DM bath, 
which have been studied in \cite{Chun:2021ief,Sen:2023uga}.

In general, the value of local dark mass $m_{\rm dark}(\mathbf{x}_\odot)$
can differ from the value in the case of vacuum mass $m_{\rm vac}$.
Due to the energy dependence of $m_{\rm dark}$ described  
by $f(E_\nu)$ 
and in particular the decrease of 
$m_{\rm dark}$ with energy below the resonance \cite{Sen:2023uga},  
the mass $m_{\rm dark}$ at $\mathcal{O}$(10\,MeV) 
energies is not restricted 
by the KATRIN experiment \cite{KATRIN:2021uub} 
and the cosmological bound \cite{PDG22-NuCos} 
where the neutrino energies are much lower. 
Therefore, $m_{\rm dark}({\bf x}_\odot)$
can be much larger than the allowed vacuum mass $m_{\rm vac}$ 
and the time delay $\Delta t_{\rm dark}$ can be enhanced. 
Interestingly, the SN1987A 
neutrino data prefer a non-zero neutrino mass 
$m_\nu\sim 3.5\,$eV, although not statistically significant \cite{Pagliaroli:2010ik}.

According to \geqn{eq:time-delay-env2}, 
the time delay depends on the local 
dark neutrino mass and the DM density profile. 
\begin{figure}[t!]
\centering
\includegraphics[width=0.9\columnwidth]{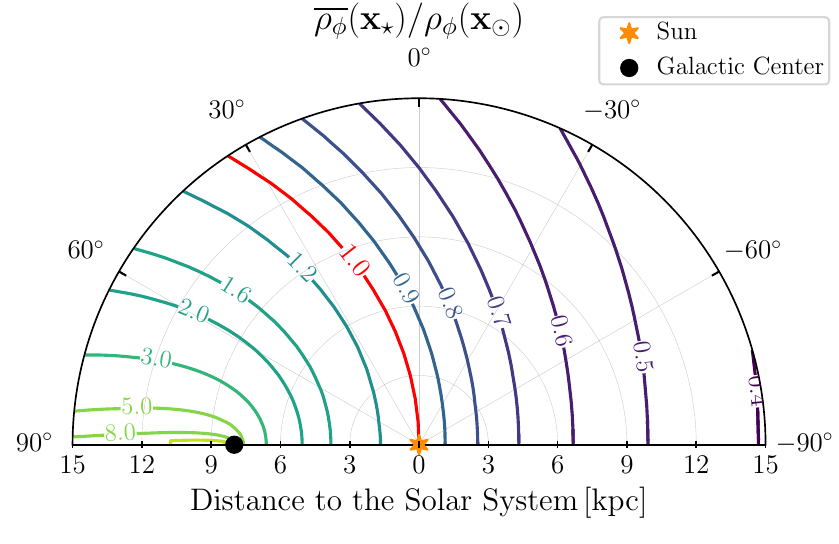}
\caption{
The lines of constant ratio $\overline{\rho_\phi}
(\mathbf{x}_\star)/\rho_\phi(\mathbf{x}_\odot)$ (numbers at the lines) 
with the Sun being at the origin in the polar coordinate system on the galactic plane. 
The polar angle of the Galactic Center (labeled by a black dot) is chosen to be $90^\circ$.   
}
\label{fig:rhobar}
\end{figure}
In our study we adopt the standard NFW profile 
\cite{Navarro:1995iw,Navarro:1996gj}, 
$ \rho_\phi(r) \equiv
0.3371/[(r/r_s)(1+r/r_s)^2]\,{\rm GeV/cm^3} $,
where $r$ is the distance from the
Galactic Center 
(GC) and $r_s =20\,$kpc is the galactic scale 
radius. The constant in the numerator 
is fixed by the DM local density,  
$\rho_\phi(r_\odot)=0.43$\,${\rm GeV/cm^3}$ 
with $r_\odot=8\,$kpc \cite{Salucci:2010qr}. 
Even for the fuzzy DM with tiny mass,
the NFW profile is a good approximation 
\cite{Ferreira:2020fam}.
Notice, however, that the DM density profile  
depends on models, see \cite{Fermi-LAT:2016afa,Young:2016ala}.
For instance, taking $D=7\,$kpc 
along the direction towards the Galactic Center 
we find that the ratio 
$\overline{\rho_\phi}(\mathbf{x}_\star)/\rho_\phi(\mathbf{x}_\odot)$
equals $3.5$, $3.3$, and $4.3$ for the NFW 
profile, the isothermal profile \cite{Bahcall:1980fb,Begeman:1991iy}, and 
the Einasto profile \cite{Graham:2005xx,Navarro:2008kc},
respectively.

\gfig{fig:rhobar} shows the values of  
$\overline{\rho_\phi} (\mathbf{x}_\star)/\rho_\phi(\mathbf{x}_\odot)$ 
on the galactic plane.  At the polar 
angle $\theta\sim 30^\circ$ (red line),
the ratio is 1, which means  
that the vacuum and dark
masses produce the same delay if the values of 
vacuum and local dark masses are equal.
The smallest relative time delay, characterized 
by the ratio $\sim 0.4$, is along the direction 
opposite to the Galactic Center, $\theta\sim -90^\circ$,  
where the nearby relative time delay varies mildly. 
On the contrary, 
the relative time delay is large and varies strongly around 
the Galactic Center due to drastic changes 
of DM density. The ratio diverges 
for neutrinos passing through
the center due to the cusp
\cite{Moore:1994yx,Flores:1994gz}. 
To avoid this divergence, we set the ratio at $\theta = 90^\circ$ to be the same as 
the one at $\theta = 89^\circ$ 
throughout this paper.
For $\theta\sim 89^\circ$ and $D\sim 9 - 11$\,kpc,
the ratio is around 11.

\textbf{SN Neutrino Flux and Delayed Event Distribution at DUNE} --
A core-collapse supernova produces neutrinos with 
$\mathcal{O}$(10\,MeV) energy.
Lasting $\mathcal{O}(10\,$s), the neutrino emission
has three distinct phases \cite{Kotake:2005zn,Scholberg:2012id}:
(i) the neutronization burst, 
(ii) the accretion phase, and (iii) the 
cooling phase. A  $\nu_e$
flux with a sharp peak in the time is 
radiated during the first phase which lasts 
$\sim 25\,$ms, while neutrinos and anti-neutrinos of 
all flavors are emitted in the other two phases.
Characteristics of the 
neutronization burst which is the main source 
of the mass sensitivity,
are rather generic and weakly depend on 
properties of a collapsing star such as the
progenitor mass, electron capture rate, as well as the equation of state
\cite{Kachelriess:2004ds,Wallace:2015xma}.

We adopt the following parametrization of 
neutrino fluxes at production introduced 
by the Garching group 
\cite{Keil:2002in,Hudepohl:2009tyy,Tamborra:2012ac,Mirizzi:2015eza}:  
\begin{equation}
  \Phi_{\nu_\beta}^0(t_{\rm e}, E_\nu)
= 
  \frac{L_{\nu_\beta}(t_{\rm e})}{\langle E_{\nu_\beta}(t_{\rm e})\rangle}
  \varphi_{\nu_\beta}(t_{\rm e}, E_\nu),
\label{eq:Phi}
\end{equation}
where $t_{\rm e}$ is the neutrino emission
time defined as the time after the shock wave bounce,  
$L_{\nu_\beta}(t_{\rm e})$ is the  
$\nu_\beta$ luminosity, 
and $\langle E_{\nu_\beta}(t_{\rm e})\rangle$ 
is the average energy of $\nu_\beta$ neutrinos. 
In \geqn{eq:Phi} $\varphi_{\nu_\beta}(t_{\rm e}, E_\nu)$ is
the $\nu_\beta$ energy spectrum 
\cite{Pompa:2022cxc}:
\begin{equation}
\hspace{-3mm}
  \varphi_{\nu_\beta}(t_{\rm e}, E_\nu)
\equiv 
  \xi_\beta(t_{\rm e})
  \left(\frac{E_\nu}{\langle E_{\nu_\beta}(t_{\rm e})\rangle}\right)^{\alpha_\beta(t_{\rm e})}
  e^{\frac{-\left(\alpha_\beta\left(t_{\rm e}\right)+1\right) E_\nu}
  {\langle E_{\nu_\beta}\left(t_{\rm e}\right)\rangle}
  },
\label{eq:varphi}
\end{equation}
with $\alpha_\beta(t_{\rm e})$ being
the pinching parameter and $\xi_\beta(t_{\rm e})$
the normalization factor determined by the condition 
$\int_0^\infty \varphi_{\nu_\beta}(t_{\rm e},E_\nu) d E_\nu =1$. 
The time dependence of parameters  
$\langle E_{\nu_\beta}(t_{\rm e})\rangle$,  
$L_{\nu_\beta}(t_{\rm e})$,
and $\alpha_\beta(t_{\rm e})$ 
is taken from the Garching Core-Collapse Modeling Group result \cite{Hudepohl:2009tyy}.

After being produced,  
neutrinos undergo adiabatic flavor conversions 
(the MSW effect) during the propagation inside the star \cite{Dighe:1999bi}.
At the star surface, the total $\nu_e$
flux emerges as 
a mixture of initially produced fluxes of different 
flavors $\Phi^0_{\nu_\beta}(t_{\rm e}, E_\nu)$: 
\begin{equation}
  \Phi_{\nu_e}(t_{\rm e}, E_\nu)
=
  p \, \Phi^0_{\nu_e}(t_{\rm e}, E_\nu)
+ (1-p)\Phi^0_{\nu_x}(t_{\rm e}, E_\nu),
\label{eq:Phi-earth}
\end{equation}
where $\Phi^0_{\nu_x}\equiv  \Phi^0_{\nu_\mu}=\Phi^0_{\nu_\tau}$. 
Here $p$ is the $\nu_e$ survival probability which equals  
$p = \sin^2\theta_{13}$ for the  normal mass ordering (NO) and  
$p = \cos^2 \theta_{13}\sin^2\theta_{12} \approx \sin^2\theta_{12}$
for inverted ordering (IO) \cite{Dighe:1999bi}. 
We neglect the collective
neutrino oscillations
which are not efficient due 
to the large flavor asymmetries during the
neutronization phase \cite{Mirizzi:2015eza}. Also, we neglect 
oscillations inside the Earth which have 
negligible impact on the neutrino mass sensitivity \cite{Pompa:2022cxc}.

\begin{figure}[t!]
\centering
\includegraphics[width=0.9\columnwidth]{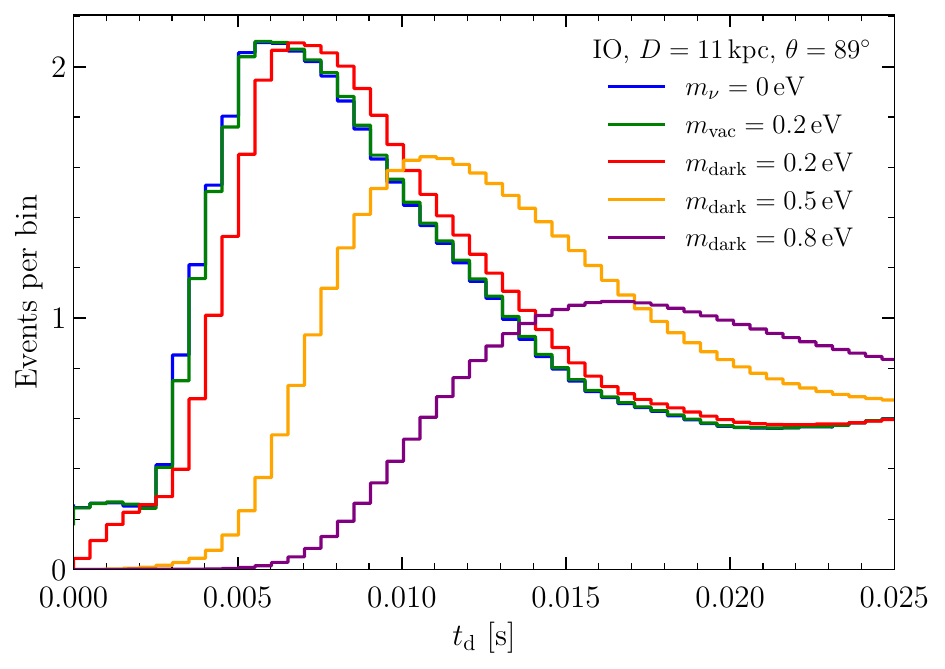}
\caption{The time 
distributions of the detected $\nu_e$
events integrated over energy for different values of vacuum and dark neutrino masses.   
}
\label{fig:SN_event}
\end{figure}

In DUNE \cite{DUNE:2015lol}, the SN neutrinos 
can be detected via the charged-current (CC) 
interaction of $\nu_e$ with Argon nuclei \cite{DUNE:2020zfm}, 
$\nu_e+^{40}{\rm Ar}\rightarrow e^-+^{40}{\rm K}^*$. 
The neutrino detection time $t_{\rm d}$
is related to its emission time $t_{\rm e}$ via,
$t_{\rm d}=t_{\rm e}+D/c+\Delta t$, where 
$D/c$ is the traveling time from  
SN to a detector for massless particles
and $\Delta t$ is the time delay due to 
the non-zero vacuum or dark   
neutrino mass. 
Since $D/c$ is a common term for all neutrinos,  
it can be dropped out,
so that $t_{\rm d} = t_{\rm e} + \Delta t$. 
Given the $\nu_e$ flux at star surface in \geqn{eq:Phi-earth},
one obtains the event rate at DUNE \cite{Pompa:2022cxc}, 
\begin{equation}
  R(t_{\rm d}, m_\nu, E_\nu) 
= 
  \frac{\Phi_{\nu_e}[t_{\rm d} - \Delta t, E_\nu]}{4\pi D^2} 
  A \sigma_{\nu_e}(E_\nu)
  \epsilon(E_\nu), 
\label{eq:event-rate}
\end{equation}
where $m_\nu = m_{\rm vac}$ or $m_{\rm dark}$ 
enters the rate via 
$\Delta t = \Delta t_{\rm vac} (E_\nu, m_{\rm vac})$ 
or $\Delta t_{\rm dark} (E_\nu, m_{\rm dark})$
as given by \geqn{eq:time-delay-usual} or \geqn{eq:time-delay-env2}. 
Here $A = 6.03\times 10^{32}$ 
is the number of Argon nuclei in the DUNE far 
detector with 40\,kt fiducial mass, 
the $\nu_e$-$^{40}{\rm Ar}$ CC 
cross section $\sigma_{\nu_e}(E_\nu)$ is
taken from the MARLEY result 
\cite{Gardiner:2020ulp,Gardiner:2021qfr},
and the detection efficiency $\epsilon(E_\nu)$
with 5\,MeV energy threshold is from \cite{DUNE:2020zfm}.

Let us consider the time delay information only that 
is given by the integral of rate in \geqn{eq:event-rate} over the neutrino energy  
$R(t_{\rm d}, m_\nu) \equiv \int d E_\nu R (t_{\rm d}, m_\nu, E_\nu)$. 
\gfig{fig:SN_event} shows the binned time distributions 
of events for SN at $D=11\,$kpc and
$\theta=89^\circ$, IO, 
and different values of vacuum and dark masses.
As the neutrino mass increases, 
the distributions are shifted to larger 
$t_{\rm d}$ and become wider with flatter front slopes.  
The difference in the distributions for 
$m_\nu = 0$\,eV, $m_{\rm vac} = 0.2$\,eV,
and $m_{\rm dark} = 0.2$\,eV
is small and therefore hard to be resolved.  
For $m_{\rm dark} \gtrsim  0.5\,$eV the
peak shifts more than 5\,ms 
which can be detected in the experiment.

We assume perfect time resolution 
since the expected DUNE time resolution   
is better than 1\,$\mu$s \cite{DUNE:2020zfm} 
which is much smaller than the neutronization
burst duration $\sim 25\,$ms.
On the other hand, the energy $E$
is further smeared by the
Gaussian function $G(E_\nu, E_{\rm d})$ with a 10\% 
fractional energy resolution 
\cite{DUNE:2020zfm} and $E_{\rm d}$ being 
the detected energy. Then the smeared 
event rate equals
\begin{equation}
\hspace{-3mm}
  R_{\rm sme}(t_{\rm d}, m_\nu,  E_{\rm d}) 
\equiv
  \int R(t_{\rm d}, m_\nu, E_\nu) G(E_\nu, E_{\rm d}) d E_\nu, 
\label{eq:r-res}
\end{equation} 
where $R(t_{\rm d}, m_\nu, E_\nu)$ is defined in \geqn{eq:event-rate}.

\textbf{Sensitivity of DUNE to Mass Origins} -- 
We consider scenarios determined by the neutrino mass ordering (MO),
the true value of the vacuum mass $m^{\rm true}_{\rm vac}$,   
and the SN coordinates $(D,\theta)$: 
$({\rm MO}, m^{\rm true}_{\rm vac}, D, \theta)$. 
For each scenario we generate a large (a few hundreds) number  
DUNE datasets ${\rm DS}_k({\rm MO}, m^{\rm true}_{\rm vac}, D, \theta)$ 
with $k$ indicating the dataset number.  
For each dataset ${\rm DS}_k$,   
using the rate $R_{\rm sme}(t_{\rm d}, m_{\rm vac}^{\rm true}, E_{\rm d})$
in \geqn{eq:r-res} with the true vacuum mass
as the target distribution we  
randomly generate $N_k$
pseudo events. 
Due to the existence of statistical 
uncertainties for a real measurement, 
$N_k$ varies between different datasets according to the 
Gaussian distribution with the mean $N \equiv \int R_{\rm sme}
(t_{\rm d}, m_{\rm vac}^{\rm true}, E_{\rm d}) dt_{\rm d} dE_{\rm d}$ and standard deviation $\sqrt{N}$.
We take $t_{\rm d} \in [0, 9]$\,s and
$E_{\rm d} \in [5, \infty]$\,MeV for the time and energy intervals, respectively.
The $i$-th pseudo event 
($i=1,...,N_{k}$) is characterized by its 
detection time as well as energy 
($t_{{\rm d},i}^k, E_{{\rm d},i}^k$)   
and sorted in the time-ascending order.

We then fit the generated pseudo events 
with two fitting parameters: 
(i) the neutrino mass $m_\nu^{\rm fit}$ (vacuum mass
$m_{\rm vac}^{\rm fit}$ or 
dark mass $m_{\rm dark}^{\rm fit}$) which enters the rate 
via the time delay $\Delta t_{i}^{{\rm fit}, k}$,
and (ii) the offset time $t_{\rm off}$ 
\cite{Pagliaroli:2008ur}. 
The time $t_{\rm off}$, being a nuisance parameter,  
accounts for the fact that the first detected event 
$t^k_{{\rm d},1}$ does not 
necessarily happen at the 
arrival time $t_0$ of the first SN 
neutrino due to the quantum nature 
of measurement:  
$t^k_{\rm off} \equiv t^k_{{\rm d},1} - t_{0} $.
Therefore, the fitting time for the $i$-th detected 
event in the $k$-th dataset equals
$
  t_{{\rm d},i}^{{\rm fit}, k}
\equiv
  t_{{\rm d},i}^k
- t_{{\rm d},1}^k
+ t^k_{\rm off}
$. 

\begin{figure}[t!]
\centering
\includegraphics[width=0.9\columnwidth]{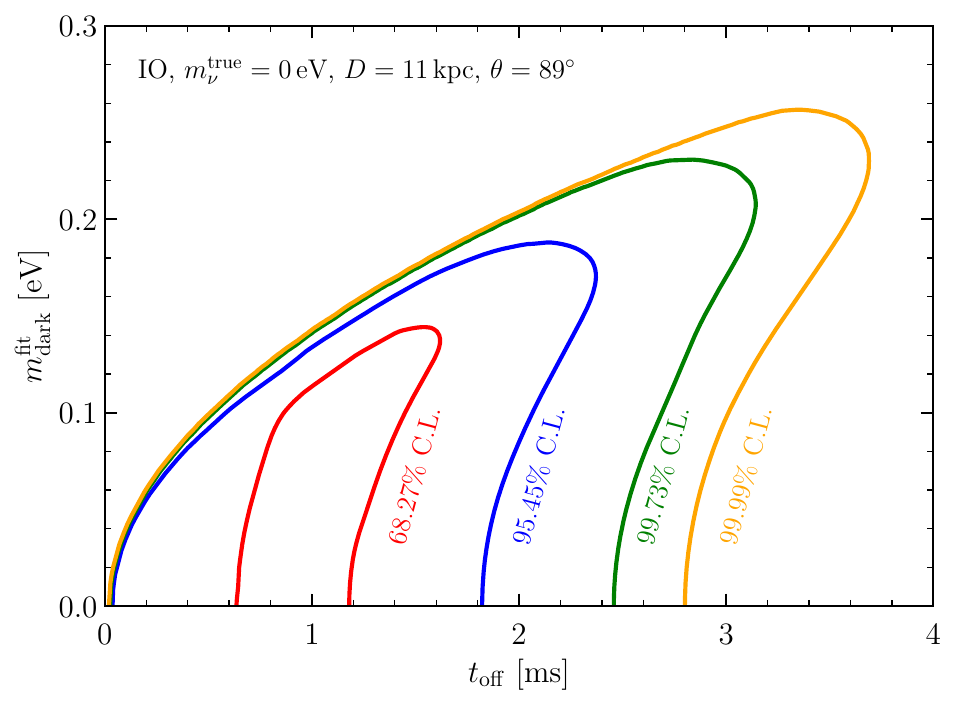}
\caption{
Contours of constant confidence levels  (numbers at the lines) 
on the $m_{\rm dark}^{\rm fit}$-$t_{\rm off}$ plane. 
}
\label{fig:SN_event2}
\end{figure}

\begin{figure*}[htbp]
\centering
\includegraphics[width=2\columnwidth]{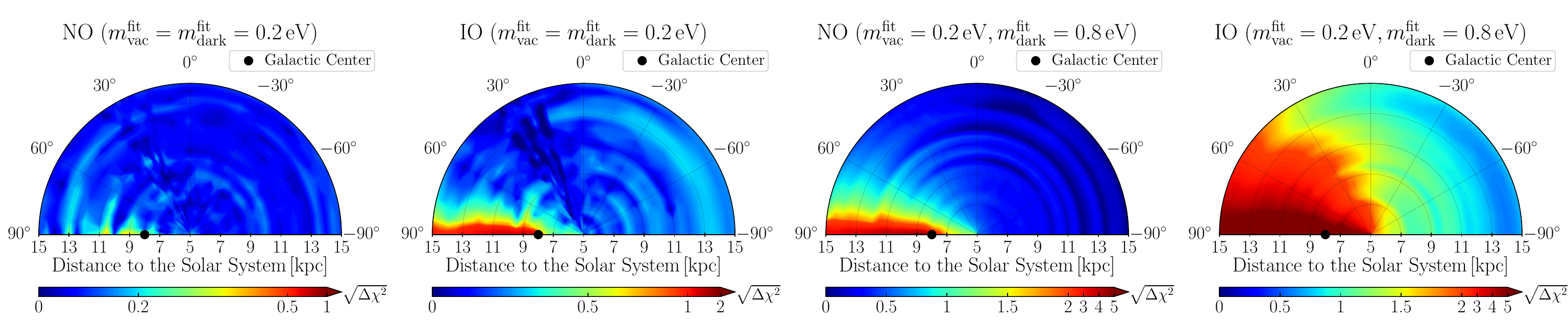}
\caption{
The difference $\sqrt{\Delta\chi^2}$   
as a function of SN location on the galactic plane for different values 
of dark mass $m_{\rm dark}^{\rm fit}$
and mass orderings.    
The central parts close to the Earth, $D < 5\,$kpc, with
small effect of dark mass variation are not shown.
}
\label{fig:SN}
\end{figure*}

We adopt the unbinned likelihood method
to evaluate the experimental sensitivities.
The likelihood $\mathcal{L}_i^k$ for the $i$-th event in dataset $k$, 
is given by its smeared detection rate function in \geqn{eq:r-res} with 
fitting mass  $m_{\rm vac}^{\rm fit}$ or
$m_{\rm dark}^{\rm fit}$ and offset time $t^k_{\rm off}$. 
Then, the total unbinned likelihood $\mathcal{L}$ 
for the $k$-th dataset is a product
\cite{Pagliaroli:2008ur,Pompa:2022cxc},
\begin{equation}
\hspace{-2mm}
  \mathcal{L}^k_{\rm tot}(m^{\rm fit}_\nu, t^k_{\rm off})
\equiv
  \prod_{i=1}^N 
  R_{\rm sme}(t^{{\rm fit}, k}_{{\rm d},i}, m_\nu^{\rm fit},  E_{{\rm d},i}^k).
\label{eq:likelihood}
\end{equation}

We first estimate the DUNE sensitivity to the 
dark mass $m_{\rm dark}$ and the offset time $t_{\rm off}$. 
Since $t^k_{\rm off}$ varies among datasets, we
consider one dataset for illustration. 
We take ${\rm MO}={\rm IO}$, $m^{\rm true}_\nu=0\,$eV, 
$D=11\,$kpc, and $\theta=89^\circ$ to generate
events. Since the mass differences are negligible
compared to the accessible mass range, we assume the 
degenerate mass spectrum 
$m^{\rm fit}_{\rm dark} \approx m_{1,2,3}$.
We then compute the total likelihood in \geqn{eq:likelihood} and consequently,  
$\chi^2_k \equiv - 2 \log \mathcal{L}^k_{\rm tot}$. 
To quantify the sensitivity to $m^{\rm fit}_{\rm dark}$ and $t_{\rm off}$,
we compute the difference  
$
\Delta \chi^2_k  \equiv \chi^2_k(m^{\rm fit}_{\rm dark}, t_{\rm off})-\chi^2_{k,{\rm min}}
$,  
where $\chi^2_{k,{\rm min}}$ is the global minimum of $\chi^2_k$ 
in the $m^{\rm fit}_{\rm dark}$-$t_{\rm off}$ space. \gfig{fig:SN_event2} displays the contours of constant confidence levels and shows that 
the DUNE experiment
is sensitive to $t_{\rm off}$ at 
the millisecond level and to $m^{\rm fit}_{\rm dark}$ below $0.2 - 0.3$\,eV. 
The best-fit values equal 
$m^{\rm fit}_{\rm dark} \sim 0\,$eV as expected,
and $t_{\rm off} \sim 1\,$ms.   
The contours show a positive correlation between 
$m_{\rm dark}^{\rm fit}$ and $t_{\rm off}$, since
these two parameters enter the likelihood 
function \geqn{eq:likelihood} in combination   
$-\Delta t^{\rm fit} + t_{\rm off} = - (m_{\rm dark}^{\rm fit}/E)^2 D/2 + t_{\rm off}$, which 
is constant at $m \sim \sqrt{t_{\rm off}}$.

Next, we study the DUNE capability to distinguish  
the dark mass from the vacuum mass for different SN locations. 
Pseudo events are generated for both mass orderings and $m^{\rm true}_{\rm vac}=0.2\,{\rm eV}$,  
which equals the KATRIN 
projected upper bound \cite{KATRIN:2021uub} and is
marginally allowed by Cosmology 
\cite{PDG22-NuCos}. We use the intervals of  distances $D\in[5, 15]\,{\rm kpc}$
and polar angles $\theta \in [-90^\circ,  90^\circ ]$.   
Recall that the dark mass can be much larger 
than the allowed vacuum mass. To test such a 
possibility,
we take the fitting vacuum mass 
$m_{\rm vac}^{\rm fit}=0.2\,{\rm eV}$ 
as well as the dark mass 
$m_{\rm dark}^{\rm fit}({\bf x}_{\odot}) = 0.2\,{\rm eV}$ 
or $0.8\,{\rm eV}$ to
compute the likelihood function in 
\geqn{eq:likelihood} and hence 
$\overline{\chi^2}(m^{\rm fit}_\nu)$
obtained by first marginalizing the 
nuisance parameter $t_{\rm off}^k$ and then
averaged over datasets.  
The difference of $\chi^2$ is computed 
between the fitting dark and vacuum masses: 
\begin{equation}
  \Delta\chi^2
\equiv 
  \overline{\chi^2}(m^{\rm fit}_{\rm dark},  D, \theta)
-
  \overline{\chi^2}(m^{\rm fit}_{\rm vac}, D).
\label{eq:Deltachi2}
\end{equation}

\gfig{fig:SN} shows the values of 
$\sqrt{\Delta \chi^2}$  
at different SN locations on the 
galactic plane. The most sensitive part appears 
for SN in the direction of the Galactic Center.  
In the case of 
$m_{\rm vac}^{\rm fit}=0.2\,{\rm eV}$ and 
$m_{\rm dark}^{\rm fit}=0.2\,{\rm eV}$ (panels 1 and 2), 
the effects of two types of masses can be 
distinguished at $\sim 2 (<1)\sigma$\,C.L. for IO (NO). 
In the case of 
$m_{\rm vac}^{\rm fit}=0.2\,{\rm eV}$ and 
$m_{\rm dark}^{\rm fit}=0.8\,{\rm eV}$ (panels 3 and 4),  the 
sensitivity for SN in the direction of the 
Galactic Center can reach $\sim 5\sigma$\,C.L. for 
both NO and IO. The sensitivity 
in the case of NO is lower than in the case of IO 
since the $\nu_e$ neutronization flux  
is highly suppressed by 
the MSW conversion as shown in \geqn{eq:Phi-earth}.

\textbf{Conclusion} --
The time delay effect of SN neutrinos opens up a unique possibility to 
test the neutrino mass origins. 
Being constant, the vacuum masses lead to
small time delay for galactic SN.
In contrast, the dark mass effects can be strongly
enhanced due to the increase of the 
DM number density and therefore the value of mass 
towards the Galactic Center.   
Therefore, observation of substantial delay effects  
would testify to the existence of the dark neutrino masses. 
Detecting SN neutrinos crossing the central regions of Galaxy 
allows to distinguish $m_{\rm vac}=0.2\,$eV 
and $m_{\rm dark} \gtrsim 0.8\,$eV for both 
mass orderings at $\sim 5 \sigma$\,C.L.
Better tests will be possible if two SN bursts from 
different locations are observed.
Further increase of sensitivity can be
done by decreasing the detection energy threshold
and analyzing events from different energy regions separately.

\textbf{Acknowledgements} --
CFK and AYS would like to thank the organizers of 
MAYORANA School \& Workshop held in July 2023,
where this project was initiated.
SFG and CFK are supported by the National Natural Science
Foundation of China (12425506, 12375101, 12090060 and 12090064) and the SJTU Double First
Class start-up fund (WF220442604).
SFG is also an affiliate member of Kavli IPMU, University of Tokyo.

\addcontentsline{toc}{section}{References}
\bibliographystyle{plain}

\end{document}